\documentclass[twoside,12pt]{article}
\usepackage{epsfig}
\usepackage{slashed}
\usepackage{amsmath}
\usepackage{amsthm}
\usepackage{amsfonts}
\usepackage{amssymb}
\usepackage{dsfont}

\def\Journal#1#2#3#4{{#1} {#2} (#4) #3 }

\def\NPB{{\em Nucl. Phys.} B}

\def\PLB{{\em Phys. Lett.} B}

\def\PRL{\em Phys. Rev. Lett.}

\def\PREP{\em Phys. Rep.}

\def\PRD{{\em Phys. Rev.} D}

\newcommand{\be}{\begin{equation}}
\newcommand{\ee}{\end{equation}}
\newcommand{\bea}{\begin{eqnarray}}
\newcommand{\eea}{\end{eqnarray}}

\begin{document}

\title{\vspace{1cm} Finite-$t$ and target mass corrections  in off-forward \\ 
hard reactions}
\author{V. M.\ Braun,$^{1}$ A. N.\ Manashov,$^{1,2}$ \\
\\
$^1$Institut f\"ur Theoretische Physik, Universit\"at
   Regensburg,\\ D-93040 Regensburg, Germany
\\
$^2$Department of Theoretical Physics,  St.-Petersburg State
University, \\
199034, St.-Petersburg, Russia
\\
}
\maketitle
\begin{abstract}
We describe a systematic approach \cite{Braun:2011zr} to the calculation of 
kinematic corrections $\propto t/Q^2, m^2/Q^2$ in hard exclusive processes 
which involve momentum transfer from the initial to the final hadron state.
As an example, the complete expression is derived for the
time-ordered product of two electromagnetic currents that
includes all kinematic corrections due to the quark distribution 
to twist-four accuracy. The results are applicable e.g. to the studies of
deeply-virtual Compton scattering.
\end{abstract}
\section{Introduction}

There is hope that hard exclusive scattering processes in Bjorken kinematics
can provide one with a three-dimensional picture of the proton in
longitudinal and transverse plane~\cite{Burkardt:2000za}, encoded in generalized parton distributions
(GPDs)~\cite{Diehl:2003ny,Belitsky:2005qn}. One of the most important reactions in this context
is Compton scattering with one real and one highly-virtual photon (DVCS) which has
received a lot of attention. The QCD description of DVCS is based on the operator product
expansion (OPE) of the time-ordered product of two electromagnetic currents.
In this language the GPDs appear as leading-twist operator matrix elements.
In order to probe the transverse proton structure one needs to measure
the dependence of the amplitude on the momentum transfer to the target
$t=(P'-P)^2$ in a broad range. Since  the available photon
virtualities $Q^2$ are limited to a few GeV$^2$ range, 
corrections of the type $\propto t/Q^2$ (which are formally higher-twist effects), 
are significant and have to be taken into account.

Such corrections are usually dubbed ``kinematic'' since they
only involve ratios of kinematic variables and at first sight have nothing to do 
with nonperturbative effects (e.g. one may consider a theoretical
limit $\Lambda_{\rm QCD}^2 \ll t \ll Q^2$). The separation of
kinematic corrections $\propto t/Q^2$ from generic twist-four corrections
$\mathcal{O}(\Lambda_{\rm QCD}^2/Q^2)$ proves, however, to be surprisingly difficult.
The problem is well known and its importance for phenomenology has been
acknowledged by many 
authors~\cite{Belitsky:2005qn,Blumlein:2000cx,Radyushkin:2000ap,Belitsky:2000vx,Kivel:2000rb,Belitsky:2001hz,Belitsky:2010jw,Geyer:2004bx,Blumlein:2006ia,Blumlein:2008di}.

The challenge is that, unlike target mass corrections in
inclusive reactions \cite{Nachtmann:1973mr}, which are determined
solely by the contributions of leading twist operators, the $\sim t/Q^2$ corrections
to off-forward processes (and for spin-1/2 targets also $\sim m^2/Q^2$ corrections)
also receive contributions from higher-twist-four operators that can be reduced
to total derivatives of the twist-two ones. Indeed,
let $\mathcal{O}_{\mu_1\ldots\mu_n}$ be a multiplicatively renormalizable
(conformal) local twist-two operator,
symmetrized and traceless over all indices. The operators
\begin{equation}
  \mathcal{O}_1 = \partial^2 \mathcal{O}_{\mu_1\ldots\mu_n}\,, \qquad
  \mathcal{O}_2 = \partial^{\mu_1}\mathcal{O}_{\mu_1\ldots\mu_n}
\label{eq:O1O2}
\end{equation}
are, on the one hand, twist-four, and on the other hand their matrix elements
are related to the leading twist matrix elements times the
momentum transfer squared (up to, possibly, target mass corrections).
Thus, both operators contribute to the  $\propto t/Q^2$, $\propto m^2/Q^2$ accuracy 
and must be taken into account.

Moreover, all these contributions are intertwined by electromagnetic gauge and 
Lorentz invariance. 
Implementation of the electromagnetic gauge invariance beyond the 
leading twist accuracy has been at the center of 
many discussions, starting from Ref.~\cite{Anikin:2000em}.
By contrast, importance of the translation invariance condition 
has never been emphasized, to the best of our knowledge.
In particular the distinction 
between the kinematic  corrections of Nachtmann's type, i.e. 
due to contributions of 
leading-twist~\cite{Radyushkin:2000ap,Belitsky:2001hz,Belitsky:2000vx,Belitsky:2010jw,%
Geyer:2004bx,Blumlein:2006ia,Blumlein:2008di}, and of higher-twist operators in 
Eq.~(\ref{eq:O1O2}) is not invariant under translations along the line connecting the 
electromagnetic currents in the $T$-product. Hence 
this distinction has no physical meaning; the existing estimates of kinematic effects, 
e.g. in DVCS, by the contributions of leading twist operators alone can be misleading.

On a more technical level, the problem arises because $\mathcal{O}_2$ has rather
peculiar properties: the divergence of a conformal operator vanishes in the free theory
(the Ferrara-Grillo-Parisi-Gatto theorem \cite{Ferrara:1972xq}).
A related feature is that using QCD equations of motion (EOM) $\mathcal{O}_2$ can be 
expressed in terms of quark-antiquark-gluon
operators. The simplest example of such a relation
is known for many years~\cite{Kolesnichenko:1984dj,Braun:2004vf,Anikin:2004ja}:
\begin{equation}
  \partial^\mu O_{\mu\nu} = 2\bar q ig G_{\nu\mu}\gamma^\mu q\,,
\label{eq:puzzle}
\end{equation}
where
$
  O_{\mu\nu} = (1/2)[\bar q \gamma_\mu\!\stackrel{\leftrightarrow}{D}_\nu \!q
  + (\mu\leftrightarrow\nu)]
$
is the quark part of the energy-momentum tensor. The operator on the r.h.s.
of Eq.~(\ref{eq:puzzle}) involves the gluon field strength and, naively,
its hadronic matrix elements are of the order of $\Lambda_{\rm QCD}^2$, which is in fact not the case.
More complicated examples can be found in \cite{Balitsky:1989ry,Ball:1998ff}.

The general structure of such relations is, schematically
\begin{equation}
 (\partial\mathcal{O})_{N} = \sum_k a^{(N)}_{k}\,{G}_{Nk}\,,
\label{eq:partialO}
\end{equation}
where ${G}_{Nk}$ are twist-four quark-antiquark-gluon operators 
and $a^{(N)}_{k}$ are the numerical coefficients.
The subscript $N$ stands for the number of
derivatives in $\mathcal{O}_N$ and the summation
goes over all contributing operators which may include total derivatives
(so that in practice $k$ is a certain multi-index).
The same operators, ${G}_{Nk}$, also appear in the OPE for
the product of currents of interest at the twist-four level:
\begin{equation}
 T\{j(x)j(0)\}^{t=4} = \sum_{N,k} c_{N,k}(x)\,{G}_{Nk}\,.
\label{eq:Tproduct}
\end{equation}
A separation of ``kinematic'' and ``dynamical'' contributions to the OPE implies
that one attempts to reassemble this expansion in such a way that the contribution
of a particular combination appearing in (\ref{eq:partialO}) is
separated from the remaining twist-four contributions.
The ``kinematic'' power correction would correspond to taking into account
this term only, and discarding contributions of ``genuine'' quark-gluon operators.

The guiding principle is that the separation of kinematic and dynamical
effects is only physically meaningful (e.g. they are separately gauge- and Lorentz-invariant)
if they have autonomous scale dependence.
Different twist-four operators of the same dimension mix with each other
and satisfy a certain renormalization group (RG) equation which can be solved,
at least in principle. Let $\mathcal{G}_{N,k}$ be the set of multiplicatively
renormalizable twist-four operators so that
\begin{equation}
  \mathcal{G}_{N,k} = \sum_{k'} \psi^{(N)}_{k,k'}\, {G}_{N,k'}\,.
\end{equation}
Eq.~(\ref{eq:partialO}) tells us that one of the solutions
of the RG equation is known {\it without  calculation}. Indeed, it provides one with
an explicit expression for a twist-four operator with the anomalous dimension 
equal to the anomalous dimension of the leading twist operator.
(For simplicity we ignore the contributions of $\partial^2\mathcal{O}_N$
in this discussion; they do not pose a problem and can
be taken into account using conventional methods.)

Let us assume that this special solution corresponds to $k=0$, i.e.
$\mathcal{G}_{N,k=0} \equiv (\partial\mathcal{O})_{N}$ and
$\psi^{(N)}_{k=0,k'}= a_{k'}$. Inverting the
matrix of coefficients, $\psi^{(N)}_{k,k'}$, and separating the term with $k=0$
we can write the expansion of an
arbitrary twist-four operator in terms of the multiplicatively renormalizable
ones
\begin{equation}
 G_{N,k} = \phi^{(N)}_{k,0} (\partial\mathcal{O})_{N} + \sum_{k'\not=0}\phi^{(N)}_{k,k'}\,
\mathcal{G}_{N,k'}\,.
\label{eq:eee}
\end{equation}
Inserting this expansion into Eq.~(\ref{eq:Tproduct}) one obtains
\begin{equation}
  T\{j(x)j(0)\}^{\rm tw-4} = \sum_{N,k} c_{N,k}(x)\phi^{(N)}_{k,0}\,(\partial\mathcal{O})_{N}
 + \ldots\,,
\label{eq:solve}
\end{equation}
where the ellipses stand for the ``genuine'' twist-four quark gluon 
operators (e.g. with different anomalous dimensions).
This is the solution we want to have, but the problem with it is that 
finding the coefficients $\phi^{(N)}_{k,0}$ in general requires knowledge of the full 
matrix $\psi^{(N)}_{k,k'}$, in other words the explicit solution
of the twist-four RG equations, which is not available.

Our starting observation is that twist-four operators in QCD come 
in two big groups: the so-called
quasipartonic~\cite{Bukhvostov:1985rn}, that only involve ``plus''
components of the fields, and non-quasipartonic which also include
``minus'' light-cone projections. 
Quasipartonic operators are not relevant for the present discussion
since they have an autonomous evolution (to one-loop accuracy).
As a consequence, $(\partial\mathcal O)_N$ does not appear in the 
expansion of quasipartonic operators in multiplicatively renormalizable ones, Eq.~(\ref{eq:eee}): 
the corresponding coefficients
$\phi^{(N)}_{k,0}$ vanish. Hence the kinematic power correction
$\sim (\partial\mathcal O)_N$ originates entirely from contributions of
non-quasipartonic operators.

Renormalization of twist-four non-quasipartonic operators was
studied recently in~\cite{Braun:2008ia,Braun:2009vc}.
The main result is that in a suitable operator basis 
the corresponding RG equations can be written in terms
of several $SL(2)$-invariant kernels. 
Using $SL(2)$-invariance we are able to prove that
the anomalous dimension matrix for non-quasipartonic operators is
hermitian with respect to a certain scalar product. This implies that
different eigenvectors are mutually orthogonal, i.e.
\begin{equation}
 \sum_k \mu^{(N)}_k \psi^{(N)}_{l,k} \psi^{(N)}_{m,k} \sim \delta_{l,m}\,,
\end{equation}
where $\mu^{(N)}_k$ is the corresponding (nontrivial) measure.
{}From this orthogonality relation and the expression (\ref{eq:partialO})
for the relevant eigenvector one obtains, for the non-quasipartonic
operators
\begin{equation}
 \phi^{(N)}_{k,0} = a^{(N)}_{k} ||a^{(N)}||^{-2}\,,
\label{eq:coc}
\end{equation}
where $||a^{(N)}||^2 =  \sum_k \mu^{(N)}_k (a^{(N)}_{k})^2$.
Inserting this expression into (\ref{eq:solve}) one ends up with 
the desired separation of kinematic effects. 

The actual derivation is done using the two-component 
spinor formalism in intermediate steps and requires some specific techniques
of the $SL(2)$ representation theory.
This talk is based on the results presented in Ref.~\cite{Braun:2011zr}; 
details of the derivation will be given in a forthcoming paper.

\section{T-product of two electromagnetic currents}

We have been able to find the contributions related to the 
leading-twist operator~(\ref{Olt}) in the $T$-product of two electromagnetic currents
$
T_{\mu\nu}=i\,T \{j_\mu^{em}(x) j_\nu^{em}(0)\}
$
to twist-four accuracy. The result can be brought to the form
\begin{align}\label{Tmn}
T_{\mu\nu} &= -\frac{1}{\pi^2x^4}\Big\{
x^{\alpha}\Big[S_{\mu\alpha\nu\beta} \mathbb{V}^\beta
+i\epsilon_{\mu\nu\alpha\beta} \mathbb{A}^\beta
\Big]
+x^2\Big[(x_\mu\partial_\nu +x_\nu\partial_\mu) \mathbb{X}
+(x_\mu\partial_\nu-x_\nu\partial_\mu) \mathbb{Y}
\Big]
\Big\}\,,
\end{align}
where $\partial_\mu = \partial/\partial x^\mu$,
$S_{\mu\alpha\nu\beta}=g_{\mu\alpha}g_{\nu\beta}+g_{\nu\alpha}g_{\mu\beta}-g_{\mu\nu}g_{\alpha\beta}$
and a totally antisymmetric tensor is defined such that $\epsilon_{0123}=1$. The
expansion of invariant functions $\mathbb{V}_\beta$ and $\mathbb{A}_\beta$ starts from  twist two,
wheareas $\mathbb{X}$ and $\mathbb{Y}$ are already twist-four.
In order to write the result we first need to introduce some notations.

We define nonlocal (light-ray) vector $O_V$ and axial-vector $O_A$  operators of the leading-twist-two
as the generating functions for local twist-two operators
\begin{align}\label{Olt}
O(z_1x,z_2x) =& \big[\bar q(z_1 x)\slashed{x}\,(\gamma_5)\, Q^2\,q(z_2 x )\big]_{l.t.}.
\end{align}
Here $x_\mu$ is an arbitrary four-vector (not necessarily light-like), $z_1$ and $z_2$ are
real numbers and $Q$ is the matrix of quark electromagnetic charges.
Here and below the Wilson line between the quark fields is implied. 
The leading-twist projector $[\ldots]_{l.t.}$ stands for the subtraction of traces 
of the local operators so that by definition
\begin{eqnarray}
\lefteqn{
\big[\bar q(z_1 x)\slashed{x}\, Q^2\,q(z_2x)\big]_{l.t.} =}
\nonumber\\
&=& \sum_{N} \frac{1}{N!} x_\mu x_{\mu_1}\ldots x_{\mu_N} \Big\{\bar q(0)\gamma_\mu
[z_1\!\stackrel{\leftarrow}{D}_{\mu_1}+z_2\!\stackrel{\rightarrow}{D}_{\mu_1}]
\ldots
[z_1\!\stackrel{\leftarrow}{D}_{\mu_N}+z_2\!\stackrel{\rightarrow}{D}_{\mu_N}]Q^2 q(0)
 -{\rm traces}\Big\}.
\end{eqnarray}
The leading-twist light-ray operators satisfy the Laplace equation
$
\partial_x^2 O(z_1x,z_2x) = 0\,.
$
The explicit form of the projector $[\ldots]_{l.t.}$  is irrelevant
for what follows. Useful representations can be found
e.g. in~\cite{Belitsky:2001hz,Balitsky:1987bk}.

Thanks to crossing symmetry the vector and axial-vector operators always appear to be antisymmetrized 
and symmetrized over the quark and antiquark positions, respectively, so we
define the corresponding combinations:
\begin{eqnarray}
{O}^{(-)}_{V}(z_1,z_2)&=\!&
\big[\bar q(z_1 x)\slashed{x}\, Q^2\,q(z_2 x )\big]_{l.t.} \!- (z_1\leftrightarrow z_2)\,,
\\
{O}^{(+)}_{A}(z_1,z_2)&=\!&
\big[\bar q(z_1 x)\slashed{x}\,\gamma_5\, Q^2\,q(z_2 x )\big]_{l.t.} \!+ (z_1\leftrightarrow z_2)\,.
\nonumber
\end{eqnarray}
The leading-twist expressions are well known and can be written as
(cf.~\cite{Balitsky:1987bk})
\begin{equation}
\mathbb{V}^{t=2}_\mu
=\frac12 \partial_\mu
\int_0^1\!{du}\,{O}^{(-)}_{V}(u,0)\,,
\qquad\qquad
\mathbb{A}^{t=2}_\mu
=\frac12\partial_\mu
\int_0^1\!{du}\,{O}^{(+)}_{A}(u,0)\,.
\end{equation}
Note that the separation of the leading-twist terms $[\ldots]_{l.t.}$ from
the nonlocal operators produces a series of kinematic power corrections
to the amplitudes, which are similar to Nachtmann target mass corrections
in deep-inelastic lepton-nucleon scattering~\cite{Nachtmann:1973mr}. Such corrections
are discussed in detail 
in~\cite{Kivel:2000rb,Belitsky:2001hz,Belitsky:2000vx,Belitsky:2010jw,Geyer:2004bx,Blumlein:2006ia,Blumlein:2008di}.

For the twist-three functions we obtain
\begin{align}\label{VV3}
\mathbb{V}^{t=3}_\mu=&
\Big[i\mathbf{P}^\nu,\int_0^1 \!\!du\,\Big\{i\epsilon_{\mu\alpha\beta\nu}
x^\alpha \partial^\beta\widetilde{{O}}^{(+)}_{A}(u)
+
\Big(S_{\mu\alpha\nu\beta}x^\alpha\partial^\beta+
\ln u \,\partial^\mu x^2\partial^\nu\Big)
\widetilde{{O}}^{(-)}_{V}(u)\Big\}\Big]\,,
\notag\\
\mathbb{A}^{t=3}_\mu=&
\Big[i\mathbf{P}^\nu,\int_0^1 \!\!du\,\Big\{i\epsilon_{\mu\alpha\beta\nu}
x^\alpha \partial^\beta\widetilde{{O}}^{(-)}_{V}(u)
+
\Big(S_{\mu\alpha\nu\beta}x^\alpha\partial^\beta+
\ln u \,\partial^\mu x^2\partial^\nu\Big)
\widetilde{{O}}^{(+)}_{A}(u)\Big\}\Big]\,.
\end{align}
Here $\mathbf{P}_\nu$ is the momentum operator
$
 [i\mathbf{P}_{\!\!\nu}, q(y)]=\frac{\partial}{\partial y^\nu} q(y),
$
and we used the notation
\begin{equation}\label{Ot}
\widetilde{{O}}^{(\pm)}_a(z)=\frac{1}{4}\int_{0}^{z}\! dw \,{O}^{(\pm)}_a(z,w)\,.
\end{equation}
One can easily verify that $x^\mu \mathbb{V}_\mu^{t=3}=\partial^\mu \mathbb{V}_\mu^{t=3}=0$
and similarly $x^\mu \mathbb{A}_\mu^{t=3}=\partial^\mu \mathbb{A}_\mu^{t=3}=0$.
Note that the terms  in $\ln u $ in Eqs.~(\ref{VV3}) are themselves twist-four 
and can be omitted if the calculation is done to twist-three accuracy. 
The resulting simplified expression is in
agreement with Refs.~\cite{Radyushkin:2000ap,Belitsky:2000vx}. These terms must be
included, however, in order to ensure correct separation of twist-three and twist-four
contributions.

The flavor-nonsinglet twist-four contributions to Eq.~(\ref{Tmn})
present our main result. In this case we prefer to write the answer in 
terms of integrals over the position of the local conformal operators, cf. Eq.~(\ref{eq:COPE}). 
This form is usually referred to as the conformal OPE~\cite{Braun:2003rp}. 
For example, a light-ray operator can be written as the conformal expansion
\begin{equation}
{O}(z_1x,z_2x) =\!\sum_N\varkappa_N\, z_{12}^N\int_0^1\! du\, (u\bar u)^{N+1}
\bigl[ \mathcal{O}_{N}(z_{12}^ux)\bigr]_{l.t.}\,,
\label{eq:COPE}
\end{equation}
where $$\varkappa_N=2(2N+3)/(N+1)!$$ and we use the shorthand notation
$ \bar u =1-u\,,\quad z_{12}=z_1-z_2\,,\quad z^u_{12} = \bar u z_1 + u z_2$.
The conformal operator $\mathcal{O}_{N}$ is defined as
\begin{align}
\mathcal{O}_{N}(y)=&
(\partial_{z_1}\!+\!\partial_{z_2})^NC_N^{3/2}\left(
\frac{\partial_{z_1}\!-\!\partial_{z_2}}{\partial_{z_1}\!+\!\partial_{z_2}}\right)
\,{O}(z_1x +y,z_2x+y)\Big|_{z_i=0},
\label{eq:On}
\end{align}
where $C_N^{3/2}(x)$ is the Gegenbauer polynomial.

The leading-twist contribution to the OPE of two electromagnetic currents 
can be written in the same form, for comparison:
\begin{equation}
 \mathbb{V}^{t=2}_\mu = 
 \partial_\mu \sum_{N,\mathrm{odd}}\frac{\varkappa_N}{N+2}
\int_0^1\!du\, u^N \bar u^{N+2}\, [\mathcal{O}^V_{N}(ux)]_{l.t.}\,.
\end{equation} 
Here $\mathcal{O}^V_{N}(ux)$ is the conformal operator
(\ref{eq:On}) at the space-time position $ux$.

We obtain
\begin{eqnarray}
\mathbb{V}^{t=4}_\mu&=&\frac12 \sum_{N,\text{odd}}\frac{\varkappa_N}{(N+2)^2}\int_0^1 du \,
\biggl\{(u\bar u)^{N+1}x_\mu\,[\widehat{\mathcal{O}}_N^V(u x)]_{l.t.}\,
\nonumber\\ &&{}\hspace*{2.5cm}
+\frac{N}{2} u^{N-1} \bar u^{N+2}\Big[u+\frac{1}{N+2}\Big]   x^2\partial_\mu\,
[(\widehat{\mathcal{O})}_N^V(u x)]_{l.t.}\biggr\}\,,
\nonumber\\
\mathbb{A}^{t=4}_\mu&=&\frac14 \sum_{N,\text{even}}\frac{\varkappa_N N}{(N+2)^2}
\int_0^1 du \, u^{N-1} \bar u^{N+2} \Big[u+\frac{1}{N+2}\Big]
x^2\partial_\mu[\widehat{\mathcal{O}}_N^A(u x)]_{l.t.}\,,
\nonumber\\
\mathbb{X}^{t=4}&=&\frac14\sum_{N,\textrm{odd}}\frac{\varkappa_N}{(N+2)^2}
\int_0^1 du\, u^{N-1}\bar u^{N+1}\Big[1-2\frac{N+1}{N+2}\bar u\Big] 
[\widehat{\mathcal{O}}_N^V(u x)]_{l.t.}\,,
\nonumber\\
\mathbb{Y}^{t=4}&=&-\frac14\sum_{N,\textrm{odd}} \frac{\varkappa_N}{(N+2)^2}
\int_0^1 du\, u^{N-1}\bar u^{N+1}\Big[1-2\frac{N+1}{N+2}\bar u
 + 2\frac{N+1}{N+3}\bar u^2\Big] 
[\widehat{\mathcal{O}}_N^V(u x)]_{l.t.}\,.
\end{eqnarray}
Here $\widehat{\mathcal{O}}_N$ is defined as
the divergence of the leading-twist conformal
operator, cf. $\mathcal{O}_2$ in~Eq.\,(\ref{eq:O1O2}):
\begin{eqnarray}
\widehat{\mathcal{O}}_N(y)&=&\frac1{N+1}\frac{\partial}{\partial x^\mu} 
\bigl[i\mathbf{P}^\mu,\mathcal{O}_N(y)\bigr]
=
\bigl[i\mathbf{P}^\mu,\mathcal{O}_{\mu\mu_1\ldots\mu_N}(y)\bigr] x^{\mu_1}\ldots x^{\mu_N}\,.
\end{eqnarray}
%
 
Note that the operator $\mathcal{O}_1$ in Eq.\,(\ref{eq:O1O2}),  which
corresponds to $[i\mathbf{P}_\mu[i\mathbf{P}^\mu, \mathcal{O}_N]$ in our present
notation, does not contribute to the answer
for our special choice of the correlation function $T\{j_\mu(x)j_\nu(0)\}$.
The T-product with symmetric positions of the currents, $T\{j_\mu(x)j_\nu(-x)\}$, 
includes both operators. The corresponding expression turns out to be much more cumbersome.

Conservation of the electromagnetic current implies that 
$\partial^\mu T_{\mu\nu}(x)=0$ and 
$\partial^\nu T_{\mu\nu}(x)=i[\mathbf{P}^\nu, T_{\mu\nu}(x)]$.
We have checked that these identities are satisfied up to twist-5 terms.

For completeness we give the relation for the operator $[i\mathbf{P}_\mu, \partial^\mu O(z_1,z_2)]$
entering the twist-three functions $\mathbb{V}^{t-3}_\mu$, $\mathbb{A}^{t-3}_\mu$
in terms of $\widehat{\mathcal{O}}_N$:
\begin{eqnarray}
[i\mathbf{P}_{\!\mu}, \partial^\mu O(z_1,z_2)]&=&
\frac12 S^+ \!\! \int_0^1 \!\!\!\!u du\, [i\mathbf{P}_\mu[i\mathbf{P}^\mu\!, O(uz_1,uz_2)]]
\nonumber\\&&{}+
\sum_{N}\!\varkappa_N(N\!+\!1)^2 z_{12}^N \!\int_0^1 \!\!\!dv\, v^{N}
\!\!\int_0^1 \!\!\!du\, (u\bar u)^{N+1} \widehat{\mathcal{O}}_N(v z_{12}^u x),
\end{eqnarray}
where $S^+=z_1^2\partial_{z_1}+z_2^2\partial_{z_2}+2z_1+2z_2$. 
It is also possible to rewrite, v.v., all contributions of local operators 
$\widehat{\mathcal{O}}_N$ in terms of the nonlocal light-ray operator 
$[i\mathbf{P}_\mu, \partial^\mu  O(z_1,z_2)]$, which can be advantageous in
certain applications.

\section{Typical matrix elements}
Hadronic matrix elements of the twist-4 operator $\widehat{\mathcal{O}}_N$ are of course related to those
of the leading twist, ${\mathcal{O}}_N$. For illustration, we present the corresponding explicit expressions
for the two proton states with momenta $p'\slashed{=} p$, which are relevant e.g. for virtual Compton scattering.
The leading-twist matrix elements can be parametrized as (cf.~\cite{Diehl:2003ny,Belitsky:2005qn})
\begin{align}
\langle p'|\mathcal{O}_N(n)|p\rangle =\bar u(p')\slashed{n}u(p)\sum_{k=even}^NF_{N,k}(t)\Delta_+^k P_+^{N-k}
+\frac1m\bar u(p')u(p)\sum_{k=even}^{N+1} H_{N,k}(t)\Delta_+^k P_+^{N+1-k}\,,
\end{align}
where $F_{N,k}(t)$ and $H_{N,k}(t)$ are generalized form factors corresponding to moments of the leading-twist 
GPD and we used the notations $P=(p+p')/2$, $\Delta = p'-p$, $p^2=(p')^2= m^2$, 
$t=\Delta^2$;  $u(p)$ is the nucleon spinor.
By analogy, we define
\begin{align}
\langle p'|\widehat{\mathcal{O}}_N(n)|p\rangle =\bar u(p')\slashed{n}u(p)\sum_{k=even}^N\widehat{F}_{N,k}(t)\Delta_+^k P_+^{N-k}
+\frac1m\bar u(p')u(p)\sum_{k=even}^{N+1} \widehat{H}_{N,k}(t)\Delta_+^k P_+^{N+1-k}\,.
\end{align}
A short calculation yields
\begin{eqnarray}
\widehat F_{N,k}(t)&=&t\,F_{N,k}(t)\frac{k(2N+3-k)}{2(N+1)^2}-\left(m^2-\frac{t}4\right)F_{N,k-2}
\frac{(N-k+2)(N-k+1)}{2(N+1)^2}
\notag\\
\widehat H_{N,k}(t)&=&t\,H_{N,k}(t)\frac{k(2N+3-k)}{2(N+1)^2}-\left(m^2-\frac{t}4\right)H_{N,k-2}
\frac{(N-k+3)(N-k+2)}{2(N+1)^2}
\nonumber\\
&&{}
-m^2 \frac{(N-k+2)}{(N+1)^2} F_{N,k-2}(t)\,.
\end{eqnarray}
Note that the twist-4 matrix elements involve both finite-$t$ and target (nucleon) 
mass corrections. Concrete applications will be considered elsewhere.

\section{Conclusions}

To summarize, we have given a complete expression for the time-ordered product 
of two electromagnetic currents that resums all kinematic corrections 
related to quark GPDs to twist-four accuracy. The results can be applied to various 
two-photon processes, e.g. to the 
studies of deeply-virtual Compton scattering and $\gamma^*\to (\pi,\eta,\ldots) +\gamma$
transition form factors. The twist-four terms
calculated in this work give rise to {\em both} a $\propto t/Q^2$ correction and 
the target mass correction $\propto m^2/Q^2$ for DVCS, 
whereas for the transition form factors these two effects
are indistinguishable as there is only one scale.
The main remaining question is whether QCD factorization itself is valid 
in such reactions to twist-four accuracy, at least for kinematic contributions. 
Clarification of this issue goes beyond the tasks of this study.

\section*{Acknowledgments}
The work by A.M. was supported by the DFG, grant BR2021/5-2,
and RFFI, grant 09-01-93108.
V.B. thanks Profs. A. Faessler and J. Wambach
for the invitation to the school and hospitality.

\end{document}